\documentclass{pasj00}

\usepackage{lscape}
\begin{document}
\SetRunningHead{Author(s) in page-head}{Running Head}
\Received{2011/02/28}
\Accepted{2011/06/24}

\title{
The Suzaku Discovery of A Hard Power-Law Component\\ in the Spectra of Short Bursts from SGR\,0501$+$4516}

\author{Yujin E. \textsc{Nakagawa}, \altaffilmark{1}
        Kazuo \textsc{Makishima}, \altaffilmark{2}
        Teruaki \textsc{Enoto}, \altaffilmark{3}}

\altaffiltext{1}{Research Institute for Science and Engineering, Waseda University, 17 Kikui-cho, Shinjuku-ku, Tokyo 162-0044}
\email{yujin@aoni.waseda.jp}
\altaffiltext{2}{Department of Physics, University of Tokyo, 7-3-1 Hongo, Bunkyo-ku, Tokyo 113-0033}
\altaffiltext{3}{Kavli Institute for Particle Astrophysiscs and Cosmology,
Department of Physics and SLAC National Accelerator Laboratory, Stanford University, Stanford, CA 94305, USA}


%

\KeyWords{stars: pulsars individual(SGR\,0501$+$4516)} 

\maketitle

\begin{abstract}
Using data with the Suzaku XIS and HXD, spectral studies of short bursts from
the soft gamma repeater SGR\,0501$+$4516 were performed.
In total, 32 bursts were detected during the $\sim$60 ks of observation conducted in
the 2008 August activity.
Excluding the strongest one, the remaining 31 bursts showed an average 2--40 keV fluence
of $1.0_{-0.5}^{+0.3}\times10^{-9}$ erg\,cm$^{-2}$.
A 1--40 keV spectrum summed over them
leaves significant positive residuals in the HXD-PIN band with $\chi^2$/d.o.f. = 74/50,
when fitted with a two-blackbody function.
By adding a power law model,
the fit became acceptable with $\chi^2$/d.o.f. = 56/48, yielding
a photon index of $\Gamma=1.0_{-0.3}^{+0.4}$.
This photon index is comparable to
$\Gamma=1.33_{-0.16}^{+0.23}$ \citep{enoto2010a}
for the persistent emission of the same object obtained with Suzaku.
The two-blackbody components showed very similar ratios, both in the temperature
and the emission radii, to those comprising the persistent emission.
However, the power-law to two-blackbody flux ratio was possibly
higher than that of the persistent emission at 2.6$\sigma$ level.
Based on these measurements, average wide-band properties of these relatively weak bursts
are compared with those of the persistent emission.
\end{abstract}

\section{Introduction}\label{intro}
Magnetars, comprising Soft Gamma Repeaters (SGRs) and Anomalous X-ray Pulsars (AXPs),
have been brought to great attention
because they are likely to have super strong surface
magnetic fields reaching $B \sim 10^{15}$ G
\citep{duncan1992, paczynski1992, thompson1995, thompson1996}.
This exceeds the critical field
strength $B_{\rm c} \equiv m_{e}^{2}c^{3}/e\hbar \approx 4.4 \times 10^{13}$ G,
where $m_{e}$, $c$, $e$, and $\hbar$ are the electron mass, the light velocity,
the electron charge and the reduced Planck constant, respectively.
To understand radiation processes in such an environment, it is necessary to
fully take into account non-perturbative effects in quantum electrodynamics.

Outstanding properties of magnetars include
burst activity, observed from all SGRs and some AXPs.
A typical ``short burst'' has a duration of $\sim$100 ms,
and a 2--100 keV energy release by 10$^{40-42}$ erg (e.g., \cite{nakagawa2007}).
Among a variety of burst activities, 
the most energetic ones are the giant flares, which were so far detected from
SGR\,0526$-$66 (e.g., \cite{mazets1979}), SGR\,1900$+$14 (e.g., \cite{hurley1999}),
and SGR\,1806$-$20 (e.g., \cite{palmer2005}).

X-ray spectra of short bursts provide useful diagnostics of their emission mechanism.
Wide-band spectra of short bursts from SGR\,1806$-$20 and SGR\,1900$+$14,
detected by High Energy Transient Explorer 2 (HETE-2; \cite{ricker2003}), are generally described by
a photoelectrically absorbed two-blackbody (2BB) model \citep{nakagawa2007},
even though this could be a phenomenological description.
Spectra of bursts \citep{feroci2004} and intermidiate flares \citep{olive2004, israel2008}
from SGR 1900$+$14 also favor the 2BB modeling.
In addition, bursts from the new magnetar SGR\,0501$+$4516,
detected by Suzaku \citep{mitsuda2007} and Swift \citep{gehrels2004},
also exhibited 2BB-type spectra
(\cite{enoto2009}, hereafter Paper\,I; \cite{kumar2010}).
In terms of this modeling, these short bursts all exhibit an interesting scaling as
$kT_{\mathrm{HT}}\sim3kT_{\mathrm{LT}}$ \citep{nakagawa2007},
where $kT_{\mathrm{HT}}$ and $kT_{\mathrm{LT}}$ are 
the higher and lower temperatures of the 2BB model, respectively.

It has long been known that SGRs and AXPs show
not only burst activities but also
persistent emission in energies below $\sim$10 keV,
of which the spectra are generally reproduced by
two phenomenological models; 2BB (e.g., \cite{tiengo2008, nakagawa2009})
or a photoelectrically absorbed blackbody plus power law model
(BB$+$PL; e.g., \cite{marsden2001, mereghetti2007}).
Moreover, recent studies of SGR\,0501$+$4516 and other objects propose
a ``blackbody plus Comptonized blackbody'' model (\cite{enoto2010a}, hereafter Paper\,II)
and a resonant cyclotron scattering model \citep{rea2008}
as alternative possibilities.
Although the spectral modeling is thus ambiguous,
the persistent X-ray emission interestingly shows the same 
$kT_{\mathrm{HT}}\sim3kT_{\mathrm{LT}}$ 
relation as those of the bursts \citep{nakagawa2009}
if we employ the 2BB representation.
This suggests a common radiation mechanism between the bursts and
persistent emission, further leading to a possibility that the persistent
X-ray emission may consist of numerous
micro bursts \citep{nakagawa2009}.

Recent studies using INTEGRAL 
(e.g., \cite{kuiper2004, molkov2005, rea2009})
and Suzaku (\cite{esposito2007, nakagawa2009}; Paper\,II; \cite{enoto2010b, enoto2010c})
revealed an extremely hard X-ray component above $\sim$10 keV
in persistent emission spectra of
a significant fraction of SGRs (including SGR\,0501$+$4516: Paper\,II) and AXPs.
The hard X-ray component, which is thought to be distinct from
the blackbody-like soft component,
can be reproduced by a power law
(PL) model with an extremely hard photon index of $\Gamma \sim 1$.
As reported in Paper\,II for SGR\,0501$+$4516 and in \citet{enoto2010c}
for some other sources, the 2--100 keV luminosity of the hard X-ray component
is often comparable to that of the soft blackbody component.
Considering these properties, as well as a clear dependence of the hardness ratio
between the hard and soft luminosities on the characteristic age as revealed with Suzaku
\citep{enoto2010c}, the hard X-ray component is expected to provide an important clue
to the nature of magnetars.
Theoretically, the hard X-ray production mechanism is extensively discussed
\citep{heyl2005, thompson2005, baring2007}, but it is not yet conclusive.

If there is a common radiation mechanism between the bursts and 
persistent emissions, the hard X-ray component
may also be seen in burst spectra.
However, short bursts of magnetars so far studied,
with fluence $>10^{-8}$ erg cm$^{-2}$, generally have
$kT_{\mathrm{HT}}=$7--15 keV (e.g., \cite{feroci2004, nakagawa2007, israel2008, kumar2010}).
As a result, their 2BB spectra, extending well up to $\gg 10$ keV, would mask any
hard X-ray component.
This raises a possibility that bursts with considerably lower fluence,
which have remained not much studied, may have lower values of $kT_{\mathrm{HT}}$,
e.g., close to those found in the persistent emission
(e.g., $kT_{\mathrm{HT}}=$0.4--3.9 keV; \cite{tiengo2008, nakagawa2009}; Paper\,II),
and would allow more sensitive searches for the hard-tail component.
Considering this, we focus on wide-band spectroscopy of low-fluence bursts.
Observations with Suzaku are suitable for this purpose, because 
of its high sensitive over a broad energy band, realized by
the X-ray Imaging Spectrometer (XIS; 0.2--12 keV; \cite{koyama2007})
and the Hard X-ray Detector (HXD; 10--600 keV; \cite{takahashi2007}).
We have hence revisited the Suzaku data of SGR\,0501$+$4516,
acquired during its 2008 August activity.
As a third publication (after Paper\,I and Paper\,II)
from this observation,
the present paper reports on our successful detection of a hard component,
in an HXD spectrum which sums over 31 short bursts from this new magnetar.

\section{Observations and Data Reductions}\label{obs}
The new soft gamma repeater SGR\,0501$+$4516
was discovered on 2008 August 22 by the burst alert telescope on-board Swift,
when it displayed SGR-like burst activity \citep{barthelmy2008}.
Soon after the discovery, a spin period of $P = 5.769\pm0.004$ was reported based
on an observation by the Rossi X-ray Timing Explorer \citep{gogus2008}.
As described in Paper\,I and Paper\,II, we triggered a Suzaku
Target-of-Opportunity (ToO) observation, which
started at 00:05 on 2008 August 26 and ended at 08:25 on 2008 August 27 (UT).
The XIS was operated with a 1/4 window option which 
yields a 2 s time resolution,
while the HXD was operated in the standard mode.
The acquired data were already utilized in Paper\,I and Paper\,II;
the former described a strong short burst and persistent soft X-ray emission, while
the latter focused on the detection of a hard component in the persistent emission.
The present paper, utilizing the same ToO data, deals with broad-band
spectra of 31 smaller short bursts.
The distance to SGR\,0501$+$4516, though estimated to be 1.5 kpc
based on its directional proximity to the young supernova remnant HB9
\citep{leathy1995, gaensler2008}, is actually very uncertain.
In this paper, the distance is hence assumed to be 4 kpc, which is
similar to the value of $\sim$5 kpc employed by \citet{rea2009}.

The reduction of the XIS and HXD event data (v2.2) 
were made using HEAsoft\,6.6.1 software.
The latest calibration database (CALDB:\,20090402)
was applied to unfiltered XIS event data using {\it xispi} (v2008-04-10).
Then, using {\it xselect} (v2.4a),
we extracted a new set of filtered XIS events
with the standard criteria\footnote{The XIS 
standard criteria were derived from http://suzaku.gsfc.nasa.gov/docs/suzaku/analysis/abc.}
and a grade selection ``GRADE = (0,2-4,6)''.
After that, hot and flickering pixels were removed using {\it cleansis} (v1.7).
Telemetry-saturated time intervals, estimated by {\it xisgtigen} (v2007-05-14),
were removed from the XIS data using {\it xselect}.
We created light curves and spectra from the cleaned XIS event data using {\it xselect}.
Response matrix files were generated by {\it xisrmfgen} (v2007-05-14),
and ancillary response function files by {\it xissimarfgen} (v2008-04-05).
The obtained net exposure is $\sim$60 ks.

Using {\it hxdpi} and {\it hxdgrade} (v2008-03-03), we applied the
latest calibration database (CALDB:\,20090902) to the unfiltered HXD event data.
Cleaned PIN and GSO events were extracted from 
these newly calibrated data with the standard
criteria\footnote{The HXD standard
criteria were taken from http://www.astro.isas.ac.jp/suzaku/analysis/7step\_HXD\_20080501.txt.}
using {\it xselect}.
Again, we created light curves and spectra
using {\it xselect}.
Dead time corrections were applied to the spectra using {\it hxddtcor}.
Response matrix files of version 2008-01-29 were used.
This yielded a net exposure of $\sim59$ ks for the HXD data.

\section{Data Analysis}\label{ana}
\subsection{Burst Detections}\label{ana_suzaku}
\begin{figure}
  \begin{center}
   \FigureFile(80mm,80mm){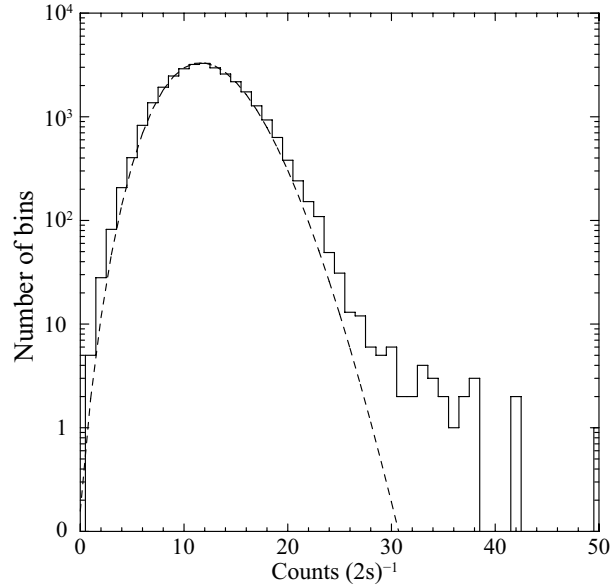}
  \end{center}
 \caption{A histogram (solid line) of the 0.2--12 keV XIS counts (per 2 s) of SGR\,0501$+$4516.
 The dashed line shows the best-fit Poisson distribution with a mean of 12.12 counts (2\,s)$^{-1}$.}\label{fig:xis_hist}
\end{figure}

As shown in figure\,1 of Paper\,I,
a number of visually obvious bursts
are found in a 0.2--12 keV XIS light curve with 2-s resolution,
obtained by summing the data from the
two FI sensors (XIS0 and XIS3) and the one BI sensor (XIS1).
At least three of them, including the strongest one analyzed in Paper\,I,
were also noticed in the 10--20 keV HXD-PIN light curve with a 500 ms time resolution.
Following our preliminary attempt in Paper\,I, we conducted a quantitative
burst search using the 0.2--12 keV light curve of the XIS.
After visually eliminating 8 obvious bursts which have $>50$ cts\,(2\,s)$^{-1}$,
the light curve was converted to a count-rate (per 2\,s) histogram as shown in figure \ref{fig:xis_hist};
this includes the background, the persistent signal emission, and short bursts. The histogram 
has an average of $\lambda=12.12\pm0.06$ and a standard deviation of $\sigma=3.48\pm0.01$,
both in units of cts\,(2\,s)$^{-1}$,
where the quoted errors refer to 90\% confidence levels,
and can be approximated by a Poissonian distribution.

\begin{figure}
  \begin{center}
   \FigureFile(160mm,80mm){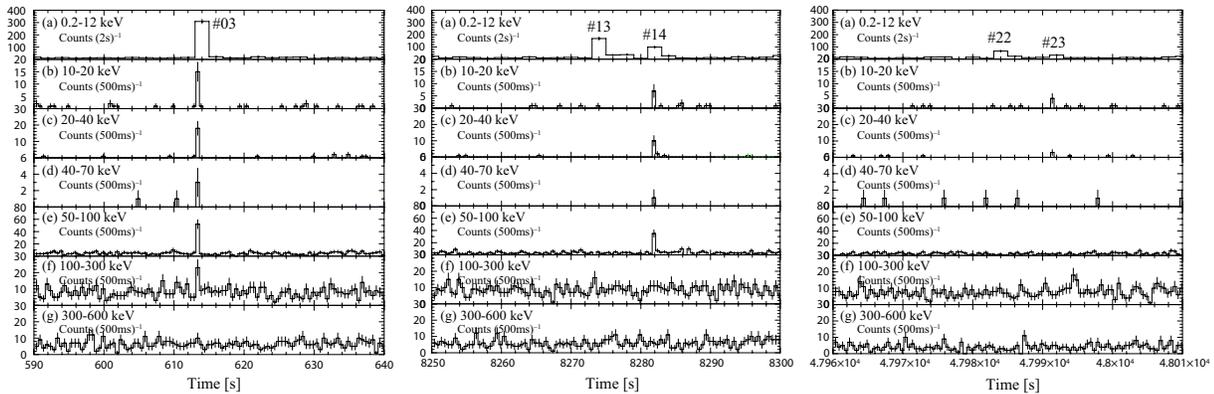}
  \end{center}
 \caption{Light curves of five typical short bursts from SGR\,0501$+$4516 recorded with the XIS
 and the HXD. The time resolution is 2\,s for the 0.2--12 keV energy band,
 and 0.5\,s for the other energy bands.
 Panels (a) are from the XIS, panels (b)-(d) are from HXD-PIN,
 while panels (e)-(g) are from HXD-GSO.}\label{fig:burst_lc}
\end{figure}

We searched the XIS count-rate histogram for those 2-s bins
where the count rate exceeds $\lambda+5\sigma \sim 30$ cts\,(2\,s)$^{-1}$.
This selection has yielded 35 time bins with significant excess counts.
Regarding a set of consecutive such bins as representing a single burst,
we thus detected 32 short bursts altogether.
Among them, the strongest one was already analyzed in Paper\,I. Below, we therefore analyze the remaining
31 bursts, which are summarized in table \ref{tab:burst_summary}.
They are hereafter identified sequentially as \#01, \#02, $\cdot\cdot\cdot$, and \#31.
These 31 short bursts are considered to be free from event pile-up effects in the XIS,
because their count rates were less than 107 cts\,(2\,s)$^{-1}$\,XIS$^{-1}$
above which the effect becomes significant
\footnote{The document is available at ftp://legacy.gsfc.nasa.gov/suzaku/nra\_info/suzaku\_td.pdf.}.
Light curves of typical short bursts (\#03, \#13, \#14, \#22 and \#23)
are presented in figure \ref{fig:burst_lc}.
Among them, two (\#03 and \#14) are accompanied by
significant emissions in the
HXD-PIN and/or HXD-GSO energy bands.

\subsection{Derivation of On-burst and Background Spectra}
Since the present paper puts its focus on burst spectra,
we must subtract the persistent emission of SGR\,0501$+$4516,
as well as the non X-ray background and the cosmic X-ray background.
For each burst, we therefore accumulate the XIS and HXD data over a time region (see below)
containing the burst, and subtract the corresponding background spectra which are acquired
before and after the burst period.
The on-burst and background data of the XIS
were both extracted from box regions
with sizes of (DETX, DETY) = ($\timeform{260''}$, $\timeform{360''}$) for XIS1,
and ($\timeform{360''}$, $\timeform{260''}$) for XIS0 and XIS3,
where DETX and DETY are detector coordinates.

Each on-burst spectrum was made using a 2-s or 6-s time interval, depending on
the burst duration in the XIS.
The corresponding background spectra were extracted from
two 10-s time intervals, one before and the other after the burst,
both separated by 2 s from the on-burst time region.
The result does not change if we instead employ 15 s for the background time intervals.
Therefore, the pulsed persistent emission
with a period of $\sim5.76$ s (Paper\,I)
does not affect the background spectrum.
If the background time intervals contained other bursts, they were eliminated
from the background spectra.

\begin{figure}
  \begin{center}
   \FigureFile(80mm,80mm){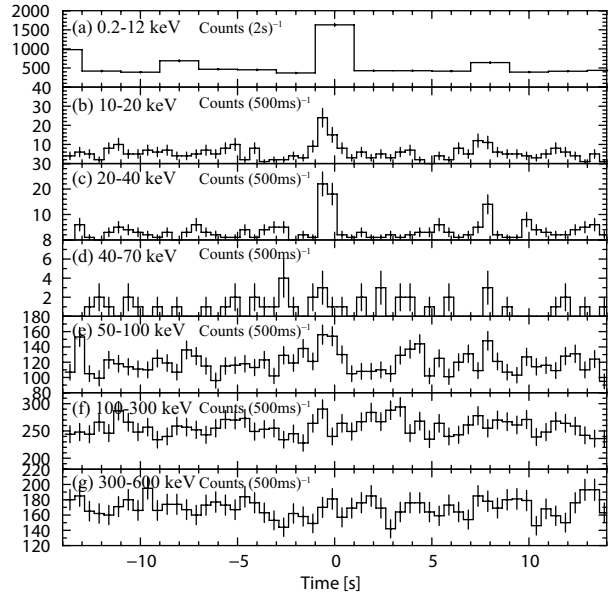}
  \end{center}
 \caption{The same as figure \ref{fig:burst_lc}, but the 31 short bursts are stacked
 referring to the XIS data.}\label{fig:lc}
\end{figure}

Figure \ref{fig:lc} shows 6-band synthetic light curves summed over
the 31 short bursts, obtained by stacking their individual light
curves in reference to the 0.2--12 keV XIS data.
Thus, the burst emission is clearly seen in the HXD-PIN data up to 40 keV,
and possibly in the 50--100 keV HXD-GSO band.
Average on-burst and background count-rates of the XIS are
$\sim$53 cts\,(2\,s)$^{-1}$ and $\sim$14 cts\,(2\,s)$^{-1}$, respectively.
Since the XIS have a time resolution of 2 s, the burst profiles in the HXD
energy bands in figure \ref{fig:lc} must be considerably smeared out.

\subsection{Spectral Analyses}\label{sec:spc_ana}
\begin{figure}
  \begin{center}
   \FigureFile(80mm,80mm){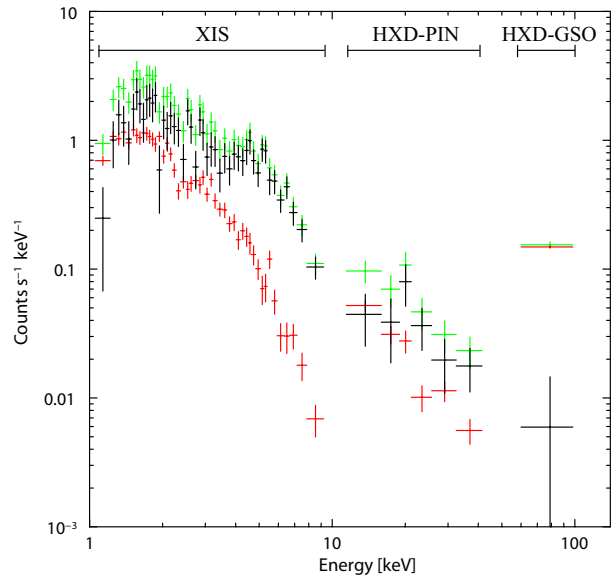}
  \end{center}
 \caption{On-burst (green), background (red), and background-subtracted (black)
 spectra of SGR\,0501$+$4516, summed over the 31 short bursts.
 The error bars represent statistical 1$\sigma$.}\label{fig:fgbg_comp}
\end{figure}

Figure \ref{fig:fgbg_comp} shows on-burst ({\it green}), background ({\it red}),
and background subtracted ({\it black}) spectra of the XIS and the HXD,
summed over the 31 bursts.
In agreement with figure \ref{fig:lc}, the burst emission is significantly detected
with HXD-PIN up to $\sim$40 keV.
Also, the burst signal may be detected marginally in the 50--100 keV GSO data.

Before quantifying the burst spectrum, let us compare it with that of the persistent
emission of SGR\,0501$+$4516.
To do this in a model-independent manner, we directly divided the summed burst
spectrum to the background-subtracted
persistent-component spectrum derived in Paper\,II.
The ratio in the XIS range was estimated using the two FI sensors.
The results, presented in figure \ref{fig:spc_ratio}, indicate that
the burst spectrum is clearly harder than the persistent emission spectrum.
In addition, the ratio in the HXD range is approximately flat,
implying that in this energy range the burst and persistent emission have
approximately the same spectral shape.

Using XSPEC 12.5.0 \citep{arnaud1996},
we fitted the summed burst spectrum with a photoelectrically absorbed 2BB model,
which has been most successful on
the short bursts from SGR\,1806$-$20 and SGR\,1900$+$14 (section \ref{intro}).
The photoelectric absorption was fixed to $8.9\times10^{21}$ cm$^{-2}$,
as estimated from the persistent X-ray emission observed by the XIS (Paper\,I).
According to a cross-calibration between the XIS and the HXD
described in \citet{kokubun2007}, the HXD normalization is
typically 13\% higher than that of the XIS for Crab data
acquired at the XIS nominal position.
Therefore the relative normalization of the HXD above the XIS was fixed to 1.13.
As shown in figure \ref{fig:spc} (b), this 2BB fits leaves significant positive residuals in energies
above $\sim$20 keV,
which makes the fit unacceptable
with $\chi^2/{\mathrm{d.o.f.}}=74/50$.
Although the 2BB fit could be improved to $\chi^2/{\mathrm{d.o.f.}}=67/49$
by allowing to vary the HXD vs. XIS relative normalizaiton,
the obtained normalization ratio, $0.2_{-0.1}^{+0.2}$,
is far outside the value of $1.13\pm0.03$ obtained from Crab
observations \citep{kokubun2007}, making the fit unrealistic.
Conversely, the 2BB fit did not improve significantly if the
HXD normalization is kept within this uncertainty range.

Given figure \ref{fig:spc_ratio}, as well as the failure of the 2BB model,
we fitted the burst spectrum with a 2BB plus power law model (2BB$+$PL).
The fit was then improved to $\chi^2/{\mathrm{d.o.f.}}=56/48$.
The PL component is considered to be significant, because an F-test indicates a
probability of $\sim$0.1\% for the fit improvement (by adding PL) to arise by chance.
The best-fit spectral parameters are summarized in table \ref{tab:spc_summary},
and the $\nu{F}_{\nu}$ form of the 2BB$+$PL fit is given in figure \ref{fig:spc} (d).
As already expected from figure \ref{fig:spc_ratio},
the power-law component indeed exhibits a photon index of
$\Gamma \sim 1$, which is comparable to that
of the hard X-ray component of the persistent emission (e.g., Paper\,II).
Consequently, we conclude that the summed short burst spectrum has a
hard-tail component,
which has never been detected in the burst spectra of any other magnetar.

Using the best-fit 2BB$+$PL spectral parameters,
a bolometric fluence of the 2BB component and a 2--40 keV fluence of the PL component
are calculated as shown in table \ref{tab:spc_summary}.
Those fluences refer to average values of the 31 short bursts,
and are lower by two orders of magnitude than a typical 2--100 keV fluence of
$\sim2\times10^{-7}$ erg\,cm$^{-2}$
for short bursts from SGR\,1806$-$20 and SGR\,1900$+$14 (e.g., \cite{feroci2004, nakagawa2007})
studied so far.
Thus, the high sensitivity of Suzaku allowed us to study, for the first time,
the wide-band properties of these low-fluence bursts.
Assuming the effective duration of the 31 short bursts from SGR\,0501$+$4516
to be 0.1 s, which is a typical value for this type of events
(e.g., \cite{feroci2004, nakagawa2007}),
we calculated the flux (luminosity)
and the blackbody radii, all averaged over
the 31 short bursts, and show the results in table \ref{tab:spc_summary}.
The effective emission radii of the two BB components,
$\sim$14 km and $\sim$1.9 km (assuming a distance of 4 kpc),
are comparable to typical values found in short bursts from
SGR\,1806$-$20 and SGR\,1900$+$14 \citep{nakagawa2007},
and from SGR\,0501$+$4516 \citep{kumar2010},
although the distance uncertainty remains.

\section{Discussion and Conclusion}\label{discussion}
Using the Suzaku ToO observation of SGR\,0501$+$4516 conducted in 2008 August,
we studied relatively dim 31 short bursts from this new magnetar.
Their average fluence, $1.0_{-0.5}^{+0.3}\times10^{-9}$ erg\,cm$^{-2}$ in 2--40 keV,
is 1--2 orders of magnitude lower than those of typical short burst studied
so far (e.g., \cite{nakagawa2007, israel2008, kumar2010}).
Following the detection of a hard component from the persistent emission of
SGR\,0501$+$4516 (Paper\,II; \cite{rea2009}), the data have allowed a
clear detection of a similar hard-tail component
in the spectrum summed over the 31 short bursts.
These results for the first time reveal spectral properties of such dim bursts, and
provide a new clue to the formation mechanisms of persistent and burst
emissions from magnetars.

As already reported in Paper\,I, the spectrum of the strongest burst (actually its precursor) from
SGR\,0501$+$4516 was well reproduced by a 2BB model,
without indication of an additional hard X-ray component.
However, this could be due to the effect mentioned in section \ref{intro}, namely,
obscuration by the high 2BB temperatures ($kT_{\mathrm{HT}}\sim14$ keV);
the data of this strong precursor are worth searching for a hard tail component.
Therefore, we re-analyzed the same pile-up and dead-time corrected spectrum 
of the precursor as studied in Paper\,I, using the 2BB$+$PL model.
The photon index was fixed to $\Gamma = 1$ to emulate the results
obtained in subsection \ref{sec:spc_ana}, and the photoelectric absorption was 
again fixed to $8.9\times10^{21}$ cm$^{-2}$ after Paper\,I.
The fit resulted in $\chi^2$/d.o.f. = 40.1/37 = 1.09, which is no better than
the value of 41.2/38 = 1.08 using the 2BB model.
Therefore, the data do not require any excess hard-tail component with $\Gamma=1.0$.
The best-fit spectral parameters are summarized in table \ref{tab:spc_summary},
which are consistent with the results in Paper\,I after renormalizing to the distance of 4 kpc
and the duration of 0.2 s.
There, the 2--100 keV flux of the PL component is given as a 90\% upper limit.

\begin{figure}
  \begin{center}
   \FigureFile(80mm,80mm){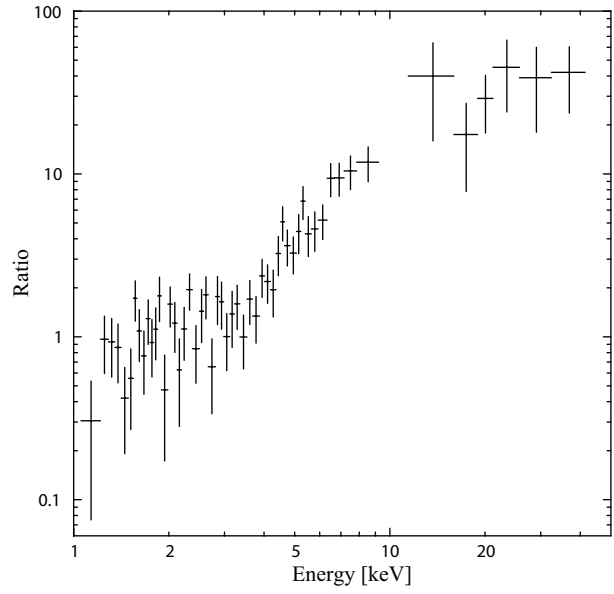}
  \end{center}
 \caption{Ratios of the spectrum summed over the 31 short bursts, to
 that of the persistent emission derived in Paper\,II. The GSO data are omitted, because
 the GSO detection of the persistent emission is not significant (Paper\,II).}\label{fig:spc_ratio}
\end{figure}

\begin{figure}
  \begin{center}
    \FigureFile(160mm,80mm){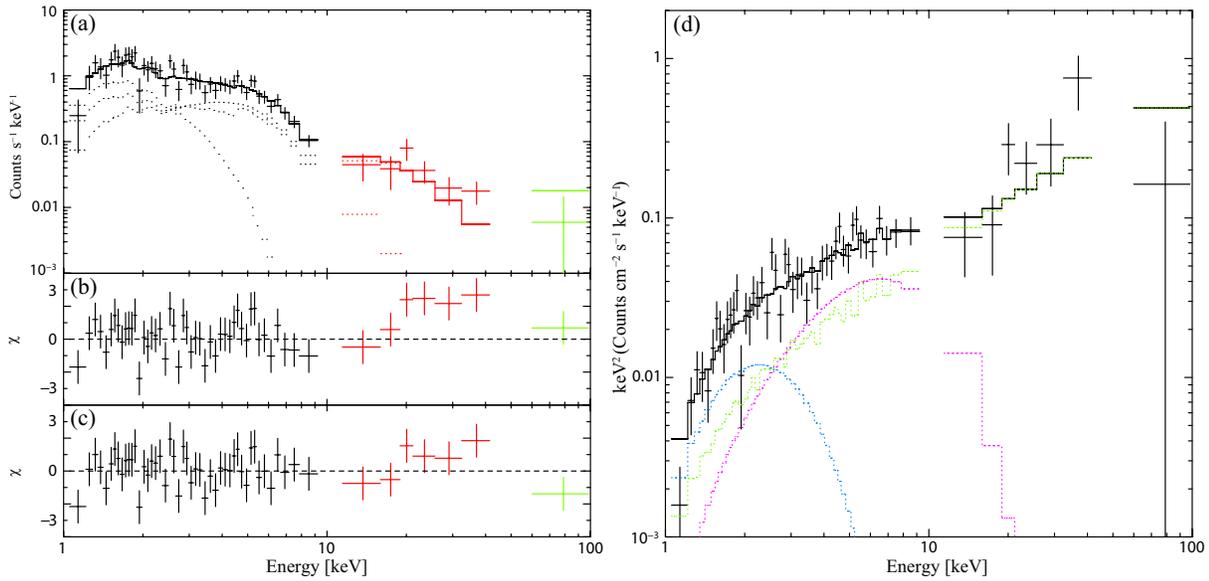}
  \end{center}
 \caption{(a) Background-subtracted XIS and HXD spectra of the summed 31 short bursts from
 SGR\,0501$+$4516, fitted jointly with a 2BB$+$PL model.
 Panels (b) and (c) show residuals from the 2BB and 2BB$+$PL fits, respectively.
 Panel (d) is the same as panel (a), but in the $\nu{F}_{\nu}$ form.}\label{fig:spc}
\end{figure}

\begin{figure}
  \begin{center}
    \FigureFile(80mm,80mm){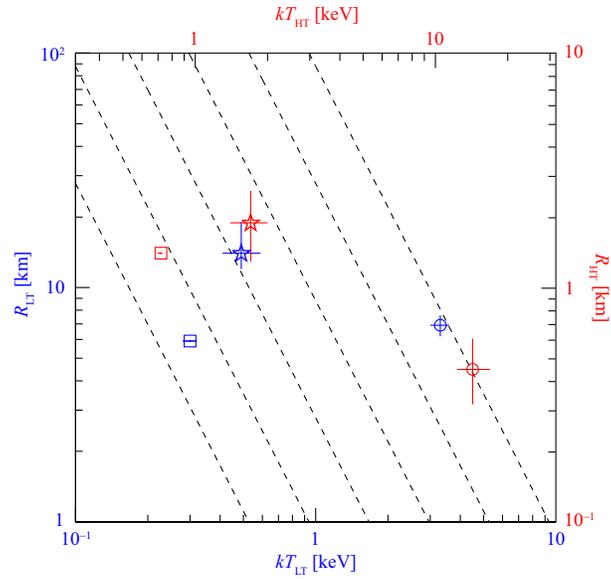}
  \end{center}
  \caption{Relations between the temperatures (abscissa) and the emission radii (ordinate) of the 2BB
  components of SGR\,0501$+$4516.
  Stars show the 31 short bursts assuming the effective duration of 0.1 s, while
  circles represent the strongest burst precursor assuming a duration of 0.2 s.
  Squares indicate the persistent emissison (Paper\,II).
  Red shows the high
  temperature component (scales at top and right), while blue the low
  temperature component (scales at bottom and left).
  Between the two components, the temperature axes are offset by a factor of $\sqrt{10}$, while the
  radius axes by a factor of 10.
  Dashed lines indicate contours of bolometric luminosity, $L$,
  which is common to the two blackbody components.}\label{fig:comp_kt_r}
\end{figure}

Given the gross spectral similarity between the short bursts
and the persistent emission of SGR\,0501$+$4516 (subsection \ref{sec:spc_ana}),
let us perform more quantitative comparison among their soft components,
referring to figure \ref{fig:comp_kt_r}
which summarizes three sets of 2BB parameters of SGR\,0501$+$4516;
the persistent emission, the 31 short bursts, and the precursor of the strongest burst.
There, we find three properties that are common to all the three spectra.
One is that  the cooler and hotter blackbodies have comparable luminosities,
and another is that $kT_{\mathrm{HT}}$ is $\sim3$ times higher than $kT_{\mathrm{LT}}$.
These 2BB properties are considered rather intrinsic to magnetars,
because they also apply to more energetic
(typically by an order of magnitude in bolometric fluence)
bursts from SGR\,1806$-$20 and SGR\,1900$+$14
observed with HETE-2 (figure 5 of \cite{nakagawa2009}),
SGR\,1900$+$14 observed with Swift \citep{israel2008},
and SGR\,0501$+$4516 observed with Swift \citep{kumar2010},
as well as to persistent emission from some other magnetars
(\cite{nakagawa2009} and references therein).
The remaining property found in figure \ref{fig:comp_kt_r} is
that the two temperatures increase with the luminosity.
In fact, the temperature of the 31 short bursts are by a factor of 4--9
lower than those of the typical bursts (e.g., \cite{feroci2004, nakagawa2007, israel2008, kumar2010}).
This justifies a posteriori our conjecture made in section \ref{intro},
i.e., a positive temperature-luminosity correlation of the 2BB component.
Incidentally, the ratio increase in figure \ref{fig:spc_ratio},
from a few keV to $\sim10$ keV, is at least partially due to the higher
2BB temperatures of the dim bursts
than those of the persistent emission.

In contrast to the present results, some published results
(e.g., \cite{kumar2010}) suggest a weak negative correlation
between the 2BB temperatures and the burst fluence.
However, these results are usually limited to rather strong bursts
with fluence $>10^{-8}$ erg cm$^{-2}$.
Then, the temperature vs. fluence might change at about this fluence.
Alternatively, weaker bursts in these studies may have actually
contained hard-tail components, and hence their spectra appeared
rather hard.

In addition to these similarities in the soft component,
the presence of a distinct PL-shaped hard component,
found in the present work, provides a novel resemblance
between the 31 short bursts and the persistent emission.
Moreover, the photon index of the former,
$\Gamma = 1.0^{+0.4}_{-0.3}$,
is consistent with the latter,  $\Gamma = 1.33^{+0.23}_{-0.16}$ (Paper\,II).
However, as already visualized by figure \ref{fig:spc_ratio},
the two phenomena can differ in their ratios
between the 2--40 keV hard-component luminosity $L_{\mathrm{PL}}$ and the bolometric
soft-component luminosity $L_{\mathrm{2BB}}$,
even excluding the effect caused by different BB temperatures.
We observed $L_{\mathrm{PL}}/L_{\mathrm{2BB}} = 2.7\pm0.9$ 
from the 31 short bursts, which is possibly higher than that of
$0.36\pm0.02$ for the persistent emission at 2.6$\sigma$ level.
Here, quoted errors are 68\% confidence levels for the ratios,
and 90\% confidence levels for the photon indices.

\begin{figure}
  \begin{center}
    \FigureFile(80mm,80mm){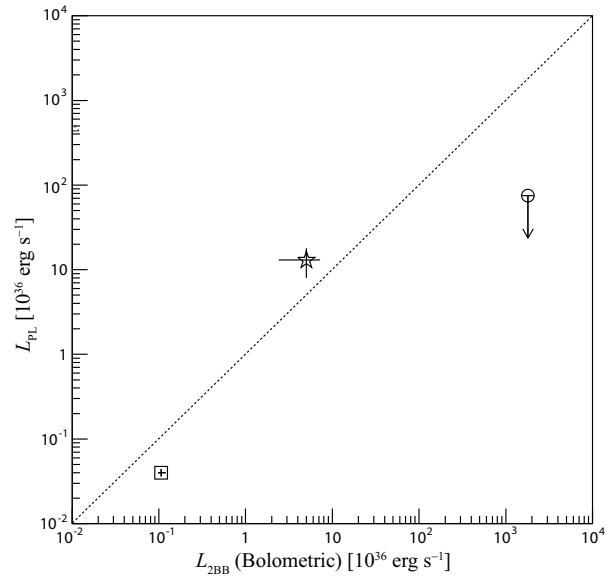}
  \end{center}
  \caption{A relation between the bolometric luminosity $L_{\mathrm{2BB}}$ of the 2BB component and the
 2--40 keV luminosity $L_{\mathrm{PL}}$ of the PL component,
 measured with Suzaku from SGR\,0501$+$4516.
 A star shows the 31 short bursts assuming the effective duration of 0.1 s,
 an open circle represents the strongest burst precursor,
 while an open square indicates the persistent emissison (Paper\,II).
 The dashed line 
 shows $L_{\mathrm{PL}}$=$L_{\mathrm{2BB}}$.
 Error bars refer to the 90\% confidence limits.}\label{fig:lbb_lpl}
\end{figure}

In order to visualize the wide-band spectral hardness,
we compare in figure \ref{fig:lbb_lpl} the relations between $L_{\rm PL}$ and $L_{\rm 2BB}$.
There, $L_{\mathrm{PL}}$ was calculated again in the 2--40 keV range.
Even though the $L_{\mathrm{PL}}/L_{\mathrm{2BB}}$ ratio could vary to some extent,
a fact of basic importance is that $L_{\mathrm{PL}}$ and $L_{\mathrm{2BB}}$ increases,
by about 2 orders of magnitude in an approximate proportion,
from the persistent emission to the dim short bursts.
This, together with the spectral similarities discussed above, suggests
that common emission mechanisms operate
between these short bursts and the persistent emission.
This in turn gives a support to our idea
that persistent X-ray emission of magnetars may in general
consist of numerous micro-bursts \citep{nakagawa2009}.
In contrast, the upper limit on the $L_{\mathrm{PL}}/L_{\mathrm{2BB}}$ ratio
obtained for the strongest burst is considerably lower
than is extrapolated from the other two less luminous data points,
suggesting some changes in the emission mechanism
from weaker to stronger short bursts.



\bigskip

This work was partially supported by Japan Society for the Promotion of Science (JSPS)
KAKENHI Grant-in-Aid for Young Scientists (B) 21740207
and Grant-in-Aid for Scientific Research (A) 22244034.



\begin{table}
 \caption{A summary of 31 short bursts from SGR\,0501$+$4516 detected by Suzaku.}\label{tab:burst_summary}
 \begin{center}
  \begin{tabular}{llr}
   \hline
   Burst & Date (UT)\footnotemark[$*$] & XIS Counts\footnotemark[$\dagger$] \\
   \hline
   \#01 & 2008-08-26 01:05:44 & 35 \\
   \#02 & 2008-08-26 01:07:02 & 30 \\
   \#03 & 2008-08-26 01:23:20 & 309 \\
   \#04 & 2008-08-26 01:24:08 & 42 \\
   \#05 & 2008-08-26 01:30:22 & 38 \\
   \#06 & 2008-08-26 02:43:44 & 34 \\
   \#07 & 2008-08-26 02:46:02 & 61 \\
   \#08 & 2008-08-26 02:46:48 & 30 \\
   \#09 & 2008-08-26 02:50:24 & 32 \\
   \#10 & 2008-08-26 02:50:38 & 36 \\
   \#11 & 2008-08-26 03:16:32 & 33 \\
   \#12 & 2008-08-26 03:16:54 & 37 \\
   \#13 & 2008-08-26 03:21:00 & 240 \\
   \#14 & 2008-08-26 03:21:08 & 99 \\
   \#15 & 2008-08-26 03:21:26 & 30 \\
   \#16 & 2008-08-26 04:38:10 & 35 \\
   \#17 & 2008-08-26 08:00:58 & 30 \\
   \#18 & 2008-08-26 09:25:06 & 31 \\
   \#19 & 2008-08-26 10:47:32 & 42 \\
   \#20 & 2008-08-26 12:45:34 & 66 \\
   \#21 & 2008-08-26 12:45:42 & 50 \\
   \#22 & 2008-08-26 14:22:50 & 66 \\
   \#23 & 2008-08-26 14:22:58 & 33 \\
   \#24 & 2008-08-26 15:27:48 & 30 \\
   \#25 & 2008-08-26 15:27:56 & 33 \\
   \#26 & 2008-08-26 15:44:14 & 32 \\
   \#27 & 2008-08-26 22:26:44 & 30 \\
   \#28 & 2008-08-27 06:07:02 & 33 \\
   \#29 & 2008-08-27 07:55:40 & 38 \\
   \#30 & 2008-08-27 07:55:48 & 31 \\
   \#31 & 2008-08-27 07:55:56 & 34 \\
   \hline
   \multicolumn{3}{@{}l@{}}{\hbox to 0pt{\parbox{180mm}{\footnotesize
   \footnotemark[$*$] {A central time of the first 2-s bin of each short burst.}
   \par\noindent
   \footnotemark[$\dagger$] Counts in a 0.2--12 keV band. The burst 13 has a 6-s duration, while the others has a 2-s duration.
   }\hss}}
  \end{tabular}
 \end{center}
\end{table}

\begin{landscape}

 \begin{table}
  \caption{Fit results of the Suzaku spectra summed over the 31 short bursts and the strongest burst precursor.\footnotemark[$*$]}\label{tab:spc_summary}
  \begin{center}
   \begin{tabular}{lllllllllll}
    \hline
    Model & $kT_{\mathrm{LT}}$\footnotemark[$\dagger$] & $R_{\mathrm{LT}}$\footnotemark[$\ddagger$] & $kT_{\mathrm{HT}}$\footnotemark[$\dagger$] & $R_{\mathrm{HT}}$\footnotemark[$\ddagger$] & $\Gamma$\footnotemark[$\S$] & $S_{\mathrm{2BB}}$\footnotemark[$\|$] & $S_{\mathrm{PL}}$\footnotemark[$\|$] & $F_{\mathrm{2BB}}$\footnotemark[$\#$] & $F_{\mathrm{PL}}$\footnotemark[$\#$] & $\chi^2$/d.o.f. \\
    & (keV) & (km) & (keV) & (km) & & & & & & \\
    \hline
    \multicolumn{11}{l}{The 31 Short Bursts} \\
    \ \ 2BB      & $0.61\pm0.14$ & ($12_{-3}^{+6}$)$t_{\mathrm{0.1}}^{-0.5}$ & $2.6_{-0.7}^{+0.9}$ & ($1.2_{-0.4}^{+0.9}$)$t_{\mathrm{0.1}}^{-0.5}$ & $\cdot\cdot\cdot$ & $0.60_{-0.10}^{+0.18}$ & $\cdot\cdot\cdot$ & ($0.60_{-0.10}^{+0.18}$)$t_{\mathrm{0.1}}^{-1}$ & $\cdot\cdot\cdot$ & 74/50 \\
    \ \ 2BB$+$PL & $0.49_{-0.08}^{+0.10}$ & ($14_{-2}^{+5}$)$t_{\mathrm{0.1}}^{-0.5}$ & $1.7\pm0.3$ & ($1.9_{-0.6}^{+0.7}$)$t_{\mathrm{0.1}}^{-0.5}$ & $1.0_{-0.3}^{+0.2}$ & $0.26\pm0.12$ & $0.7\pm0.3$ & ($0.26\pm0.12$)$t_{\mathrm{0.1}}^{-1}$ & ($0.7\pm0.3$)$t_{\mathrm{0.1}}^{-1}$ &  56/48 \\
    \multicolumn{11}{l}{The Strongest Burst Precursor} \\
    \ \ 2BB$+$PL & $3.3\pm0.3$ & $6.9\pm0.7$ & $14.3_{-2.0}^{+2.6}$ & $0.45_{-0.13}^{+0.16}$ & 1.0 (fixed) & $190\pm30$ & $<20$ & $100\pm20$ & $<10$ & 40/36 \\
    \hline
    \multicolumn{11}{@{}l@{}}{\hbox to 0pt{\parbox{230mm}{\footnotesize
      \footnotemark[$*$] The precursor duration of 0.2 s is assumed. The photoelectric absorption is fixed to $8.9\times10^{21}$ cm$^{-2}$.
      The distance is assumed to be 4 kpc. The quoted errors are 90\% confidence levels.
      \par\noindent
      \footnotemark[$\dagger$] $kT_{\mathrm{LT}}$ and $kT_{\mathrm{HT}}$ denote blackbody temperatures.
      \par\noindent
       \footnotemark[$\ddagger$] $R_{\mathrm{LT}}$ and $R_{\mathrm{HT}}$ denote the emission radii, with $t_{\mathrm{0.1}}$ the burst duration in 0.1 s unit.
      \par\noindent
      \footnotemark[$\S$] $\Gamma$ denotes the photon index of the PL component.
      \par\noindent
      \footnotemark[$\|$] $S_{\mathrm{2BB}}$ and $S_{\mathrm{PL}}$ denote bolometric and 2--40 keV (2--100 keV for the precursor) fluence of
      the 2BB and PL components, respectively, in units of $10^{-9}$ erg\,cm$^{-2}$.
      \par\noindent
      \footnotemark[$\#$] $F_{\mathrm{2BB}}$ and $F_{\mathrm{PL}}$ are fluxes corresponding to
      $S_{\mathrm{2BB}}$ and $S_{\mathrm{PL}}$ respectively, in units of $10^{-8}$ erg\,cm$^{-2}$\,s$^{-1}$.
    }\hss}}
   \end{tabular}
  \end{center}
 \end{table}
\end{landscape}

\onecolumn

\begin{thebibliography}{}
 \bibitem[Arnaud(1996)]{arnaud1996} Arnaud, K. A. 1996, in Astronomical Data Analysis Software and Systems V,
			eds. G. Jacoby and J. Barnes, ASP Conference Series, 101, 17
 \bibitem[Baring \& Harding(2007)]{baring2007} Baring,~M.~G., \& Harding,~A.~K. 2007, \apss, 308, 109
 \bibitem[Barthelmy et al.(2008)]{barthelmy2008} Barthelmy,~S.~D. \etal\ 2008, GRB Coord. Netw. Circ., 8113
 \bibitem[Duncan \& Thompson(1992)]{duncan1992} Duncan,~R., \& Thompson,~C. 1992, \apj, 392, L9
 \bibitem[Enoto(2010)]{enoto2010phd} Enoto,~T. 2010, Ph.D. Thesis, University of Tokyo
 \bibitem[Enoto et al.(2009)]{enoto2009} Enoto,~T. \etal\ 2009, \apj, 693, L122
 \bibitem[Enoto et al.(2010a)]{enoto2010a} Enoto,~T. \etal\ 2010a, \apj, 715, 665
 \bibitem[Enoto et al.(2010b)]{enoto2010b} Enoto,~T. \etal\ 2010b, \pasj, 62, 475
 \bibitem[Enoto et al.(2010c)]{enoto2010c} Enoto,~T. \etal\ 2010c, \apj, 722, L162
 \bibitem[Esposito et al.(2007)]{esposito2007} Esposito,~P. \etal\ 2007, \aap, 476, 321
 \bibitem[Feroci et al.(2004)]{feroci2004} Feroci,~M., Calliandro,~G.~A., Massaro,~E.,
			Mereghetti,~S., \& Woods,~P.~M. 2004, \apj, 612, 408
 \bibitem[Gaensler et al.(2008)]{gaensler2008} Gaensler,~B.~M., \etal\ 2008, GRB Coord. Netw. Circ., 8149
 \bibitem[G\"{o}\u{g}\"{u}\c{s} et al.(2008)]{gogus2008} G\"{o}\u{g}\"{u}\c{s},~E., Woods,~P., \& Kouveliotou,~C. 2008,
			GRB Coord. Netw. Circ., 8118
 \bibitem[Gehrels et al.(2004)]{gehrels2004} Gehrels,~N., \etal\ 2004, \apj, 611, 1005
 \bibitem[Heyl \& Hernquist(2005)]{heyl2005} Heyl,~J.~S., \& Hernquist,~L. 2005, \mnras, 362, 777	
 \bibitem[Hurley et al.(1999b)]{hurley1999} Hurley,~K., \etal\ 1999, \nat, 397, 41
 \bibitem[Israel et al.(2008)]{israel2008} Israel,~G.~L., \etal\ 2008, \apj, 685, 1114
 \bibitem[Kokubun et al.(2007)]{kokubun2007} Kokubun,~M., \etal\ 2007, \pasj, 59, 53
 \bibitem[Koyama et al.(2007)]{koyama2007} Koyama,~K., \etal\ 2007, \pasj, 59, S23
 \bibitem[Kuiper, Hermsen, Mendez(2004)]{kuiper2004} Kuiper,~L., Hermsen,~W., \& Mendez,~M. 2004, \apj, 613, 1173
 \bibitem[Kumar, Ibrahim, Safi-Harb(2010)]{kumar2010} Kumar,~H.~S., Ibrahim,~A.~I., \& Safi-Harb,~S. 2010, \apj, 716, 97
 \bibitem[Leathy \& Aschenbach(1995)]{leathy1995} Leathy,~D.~A., \& Aschenbach,~B. 1995, \aap, 293, 853
 \bibitem[Marsden \& White(2001)]{marsden2001} Marsden,~D., \& White,~N.~E. 2001, \apj, 551, L155
 \bibitem[Mazets, Golenetskii \& Gur'yan(1979)]{mazets1979} Mazets,~E.~P., Golenetskii,~S.~V.,
			\& Gur'yan,~Yu.~A. 1979, Soviet Astron. Lett., 5, 343
 \bibitem[Mereghetti, Esposito \& Tiengo(2007)]{mereghetti2007} Mereghetti,~S., Esposito,~P., \& Tiengo,~A. 2007, \apss, 308, 13
 \bibitem[Mitsuda et al.(2007)]{mitsuda2007} Mitsuda,~K., \etal\ 2007, \pasj, 59, S1
 \bibitem[Molkov et al.(2005)]{molkov2005} Molkov,~S., Hurley,~K., Sunyaev,~R., Shtykovsky,~P., Revnivtsev,~M.,
			\& Kouveliotou,~C. 2005, \aap, 433, L13
 \bibitem[Nakagawa et al.(2007)]{nakagawa2007} Nakagawa,~Y.~E., \etal\ 2007, \pasj, 59, 653
 \bibitem[Nakagawa et al.(2008)]{nakagawa2008} Nakagawa,~Y.~E., \etal\ 2008, GRB Coord. Netw. Circ., 8265
 \bibitem[Nakagawa et al.(2009)]{nakagawa2009} Nakagawa,~Y.~E., \etal\ 2009, \pasj, 61, S387
 \bibitem[Olive et al.(2004)]{olive2004} Olive,~J.~-F. \etal\ 2004, \apj, 616, 1148
 \bibitem[Paczy\'{n}ski(1992)]{paczynski1992} Paczy\'{n}ski,~B. 1992, Acta Astron., 42, 145
 \bibitem[Palmer et al.(2005)]{palmer2005} Palmer,~D.~M., \etal\ 2005, \nat, 434, 1107
 \bibitem[Rea et al.(2008)]{rea2008} Rea,~N., Zane,~S., Turolla,~R., Lyutikov,~M., G\"{o}tz,~D. 2008, \apj, 686, 1245
 \bibitem[Rea et al.(2009)]{rea2009} Rea,~N., \etal\ 2009, \mnras, 396, 2419
 \bibitem[Ricker et al.(2003)]{ricker2003} Ricker,~G., \etal\ 2003, in Gamma-Ray Bursts
			and Afterglow Astronomy, ed. G.~R.~Ricker \& R.~Vanderspek (Melville: AIP), 662, 3
 \bibitem[Takahashi et al.(2007)]{takahashi2007} Takahashi,~T., \etal\ 2007, \pasj, 59, S35
 \bibitem[Tanaka et al.(2008)]{tanaka2008} Tanaka,~Y.~T., Terasawa,~T., Yoshida,~M., Horie,~T., Hayakawa,~M. 2008, \jgr, 113, A07307
 \bibitem[Thompson \& Beloborodov(2005)]{thompson2005} Thompson,~C., \& Beloborodov,~A.~M. 2005, \apj, 634, 565
 \bibitem[Thompson \& Duncan(1995)]{thompson1995} Thompson,~C., \& Duncan,~R. 1995, \mnras, 275, 255
 \bibitem[Thompson \& Duncan(1996)]{thompson1996} Thompson,~C., \& Duncan,~R. 1996, \apj, 473, 322
 \bibitem[Tiengo et al.(2008)]{tiengo2008} Tiengo,~A., Esposito,~P., \& Mereghetti,~S. 2008, \apj, 680, 133
\end{thebibliography}
\end{document}